\documentclass[]{aa}
\usepackage{graphicx}
\begin{document}
%---------- page titre-----------------------
\title{Constraints on the disk geometry of the T Tauri 
star AA~Tau from linear polarimetry.\thanks{Based on
observations collected with the {\sc sterenn} Polarimeter at the 2m
Bernard-Lyot telescope (TBL) operated by INSU/CNRS and Pic-du-Midi
Observatory (CNRS USR 5026), France.}}
%
%-- liste des auteurs ----
%
\author{
 Fran\c cois M\'enard \inst{1}
\and
 J\'er\^ome Bouvier  \inst{1} 
\and
 Catherine Dougados  \inst{1}         
\and
 Stanislav Y. Mel'nikov \inst{2,3}
\and
 Konstantin N. Grankin \inst{2,3}
}
%-------   
\offprints{F.M\'enard}
% ----- liste des adresses ----------
%
\institute{  
Laboratoire d'Astrophysique de Grenoble, CNRS/UJF UMR~5571, 414 rue de
la Piscine,\\ BP 53, F-38041 Grenoble cedex~9, France\\
  \email{menard@obs.ujf-grenoble.fr; jbouvier@obs.ujf-grenoble.fr; dougados@obs.ujf-grenoble.fr}
\and  
Astronomical Institute of the Academy of Sciences of
Uzbekistan, Astronomicheskaya 33, Tashkent 700052,\\ Uzbekistan\\
\email{kn@astrin.uzsci.net; smeln@sai.msu.ru}
\and
Isaac Newton Institute of Chile, Uzbekistan Branch, Santiago, Chile\\
\email{newton@reuna.cl}
}

\date{Received ; accepted}

\abstract{
We have simultaneously monitored the photometric and polarimetric
variations of the Classical T Tauri star AA~Tau during the fall of
2002. We combine these data with previously published polarimetric
data covering two earlier epochs. The phase coverage is complete,
although not contiguous. AA~Tau clearly shows cyclic variations
coupled with the rotation of the system. The star-disk system produces
a repeatable polarisation curve where the polarisation increases with
decreasing brightness.  The data fit well with the model put forward
by Bouvier et al. (1999) where AA~Tau is viewed almost edge-on and its
disk is actively dumping material onto the central star via
magnetospheric accretion. The inner edge of the disk is deformed by
its interaction with the tilted magnetosphere, producing ``eclipses''
as it rotates and occults the photosphere periodically. From the shape
of the polarisation curve in the QU-Plane we confirm that the
accretion disk is seen at a large inclination, almost edge-on, and
predict that its position angle is PA $\sim90^o$, i.e., that the
disk's major axis is oriented in the East-West direction.
}
\authorrunning{M\'enard et al.}
\titlerunning{Polarimetric monitoring of AA~Tau}

   \maketitle
%
% ------------- fin de page titre -----------------
%
\section{Introduction}

Classical T Tauri stars (CTTS) are low-mass pre-main sequence stars
surrounded by accretion disks. The current models for these star-disk
systems propose that the disks are truncated at a few stellar radii by
a powerful stellar magnetic field. Accretion onto the central star is
proceeding along the field lines, from the inner edge of the disk to
the star, at near free-fall velocities. It is the so-called
magnetospheric accretion model. Observational evidence for this
scenario has been growing steadily, e.g., Bertout et al. (1988),
Edwards et al. (1994), Alencar, Johns-Krull \& Basri (2001).
  
One of the stringent assumptions of this model is the topology and
strength of the stellar magnetic field required. It is usually assumed
to be dipolar with a strength of the order of a few kiloGauss or
so. Apart from a few observations (e.g., Guenther et al. 1999;
Johns-Krull et al. 1999, 2001; Donati et al. 2000), its exact nature
remains poorly constrained. A few authors have argued that the field
may not be aligned with the stellar rotation axis in a few cases
(e.g., Kenyon et al. 1994 for DR~Tau; Johns \& Basri 1995 for SU~Aur;
Bouvier et al. 1999 for AA~Tau).

To probe further into the details of magnetospheric accretion, we have
obtained spectro-photometric synoptic observations of a few
representative CTTS, including AA~Tau (Bouvier et al. 1999, hereafter
B99). Continuous coverage over several rotation periods provided a
detailed view of the star-disk interaction, hence of the physics of
the accretion process itself.

Specific to AA~Tau, B99 proposed a model where the
star-disk system is seen almost edge-on and where the inner edge of
the disk is deformed by its interaction with the magnetic field. As a
consequence and as the disk rotates, occulting material moves in and
out of the line-of-sight, occulting the star and producing quasi-cyclic
fadings. Because the dimming episodes are not strictly identical, even
from one cycle to the next, the accretion process is shown to be
non-steady, with the mass-loading of the field lines and the
deformation of the disk highly time-dependent (Bouvier et al. 2003).

All these phenomena occur well beyond the resolution limit of current
telescopes and one will have to wait for the next generation of
long-baseline optical/near-infrared interferometers for any hope of
directly imaging these regions. Until then, ``tomography'' as obtained
by synoptic campaigns is our best tool to understand the details of
the linkage between the stellar magnetosphere and its accretion disk
and therefore to understand the fundamental mass-accretion process by
which solar-like stars form.

In the following sections we present photometric and linear
polarimetric monitoring of AA~Tau in an attempt to solve some of the
issues raised by the model proposed by B99. First,
is it really circumstellar extinction that causes the quasi-periodic
fadings?  Second, what is the 3-dimensional orientation of the disk,
is it really edge-on?

Understanding these questions goes beyond understanding AA~Tau itself.
It will help demonstrate whether or not the accretion topology derived
for AA~Tau can be applied more generally to other CTTS.

%----------------------------------------
\section{Observations and Data Reduction}
\subsection{Linear Polarimetry}
\label{sec:obspol}
Three data sets are presented in Table~\ref{tab:poldata}.  Data sets 1
\& 2 were published previously by M\'enard \& Bastien (1992) and 
B99, respectively. Data set 3 is new.  It was
obtained with the {\sc sterenn} polarimeter attached to the 2m
Bernard-Lyot Telescope of the Pic-du-Midi Observatory, France. The
observations were performed during the period 30 October -- 5 November
2002\footnote{The same instrument was used to collect some of the data
in set 2, but the detectors were different.}.

The data reduction procedure for data sets 1 \& 2 is presented in the
original papers.  In the fall of 2002, for data set 3, the efficiency
and stability of the instrument was checked every night by measuring at
least 2 of a set of 4 highly polarised standard stars. Night-to-night
efficiency corrections were applied to account for small variations in
the instrument efficiency. The average efficiency of the instrument
was 0.95. A constant shift of $+1.25^o$ in position angle (i.e.,
obs. value + 1.25 = true value), induced by the chromaticity of the
rotating half-wave plate was carefully measured and removed from the
data.  We found no instrumental polarisation by measuring well known
unpolarised standard stars and numerous nearby M-dwarfs during
moon-free time. With a typical 1-$\sigma$ error of $\sigma {\rm(
P)}$=0.02\%, all these measurements are compatible with zero
polarisation and no correction was applied to the data. 

\begin{table}[htb]
\centering
\caption{Linear polarisation measurements of AA~Tau}
\begin{tabular}{llllll}
\hline
JD & filter & P & $\sigma$(P) & $\theta$ & $\sigma(\theta)$ \\ 
   & &(\%)& (\%) & ($^o$) & ($^o$) \\
\hline
\noalign{\medskip Data set 1: ~~7 -- 18/10/1984 \hfill \smallskip}
 2445980.952 & 4700 & 0.73  & 0.07  & 43.3 & 2.5 \\ 
 2445982.862 & 4700 & 2.14  & 0.11  & 12.6 & 1.5 \\ 
 2445983.914 & 4700 & 2.35  & 0.21  & 11.3 & 2.5 \\
 2445991.801 & 4700 & 1.68  & 0.23  & 12.0 & 3.9 \\
 2445991.840 & 4700 & 1.75  & 0.22  &  6.4 & 3.5 \\ 
\noalign{\medskip Data set 2: ~~23 -- 29/11/1995 \hfill  \smallskip}
 2450042.823 & V    & 0.72 & 0.01 & 25.0 & 1.0 \\ 
 2450043.979 & V    & 0.70 & 0.01 & 24.5 & 1.0 \\
 2450044.891 & V    & 1.08 & 0.01 & 14.8 & 1.0 \\
 2450048.547 & V    & 1.33 & 0.13 &  2.2 & 2.4 \\
 2450049.525 & V    & 0.60 & 0.09 & 20.3 & 3.7 \\
 2450049.676 & V    & 0.80 & 0.21 & 20.8 & 6.5 \\
 2450050.638 & V    & 0.84 & 0.10 & 27.5 & 2.9 \\ 
\noalign{\medskip Data set 3: ~~30/10 -- 5/11/2002 \hfill \smallskip}
%
% raw data
 2452577.586 & I    & 0.80 & 0.05 & 27.8 & 2.0 \\ 
 2452579.520 & I    & 2.11 & 0.06 &  7.0 & 1.1 \\ 
 2452580.544 & I    & 0.54 & 0.06 & 37.0 & 3.1 \\
 2452583.510 & I    & 0.47 & 0.03 & 34.0 & 2.0 \\
 2452584.651 & I    & 0.64 & 0.04 & 19.0 & 1.9 \\
\hline
\end{tabular}
\label{tab:poldata}
\end{table}
%
% set1 -> from Menard et Bastien 1992
% set2 -> from Bouvier et al. 1999 (Schmidt KPNO  + Menard PDM)
% set3 -> PDM , novembre 2002

In Table~\ref{tab:poldata}, the julian date of the middle of the
observation is given for each datum. The next column gives the central
wavelength of the filter or the photometric band used. Then follow the
polarisation, P, its associated error, $\sigma$(P), the polarisation
position angle given in the equatorial system, $\theta$, and its
associated error, $\sigma$($\theta$). The error $\sigma$($\theta$) was
increased to 1.0$^o$ systematically in cases where it would have been
smaller formally.

\subsection{Optical Photometry}

The optical aperture photometry used here was obtained at the
60cm telescope of the Mount Maidanak Observatory, Uzbekistan, during
the period 13 September - 1 December 2002. A photometer with a pulse
counting FEU-79 photomultiplier tube was used with standard UBVR
Johnson filters and a 15$''$ diaphragm. Typical exposure times ranged
from 50 up to 120 seconds, depending on the filter and on the
brightness of the star. During one of the photometric nights of the
run, secondary standards were observed for flux calibration. The data
were reduced with the standard procedures using the mean extinction
coefficients for the observatory. The final observing error is
$\sim$0.01 mag. on average. In this paper we present only the V-band
data overlapping with the polarisation observations. All other data
will be made available elsewhere.

AA~Tau's photometry has also been monitored prior to this work
and rotational periods of roughly 8.2 days have been reported, e.g.,
Vrba et al. (1989) and Herbst et al. (1994). B99 presented the results
of a large synoptic campaign performed in 1995, matching with data set
2 of Table~\ref{tab:poldata}. In that campaign, the 8.2d rotational
period was recovered and quasi-cyclic episodes of fadings were found
due to the improved time coverage, see also Fig.~\ref{fig:pvsv-95}. 
The behaviour of AA~Tau in the fall of 2002, i.e., in the data 
presented here, appears very similar, see Fig.~\ref{fig:pvsv-02}.

%----------------------------------------
\section{Results}

\subsection{Linear Polarimetry}

\begin{figure}[ht]
\centering
\includegraphics[width=4.7cm, angle=-90.0]{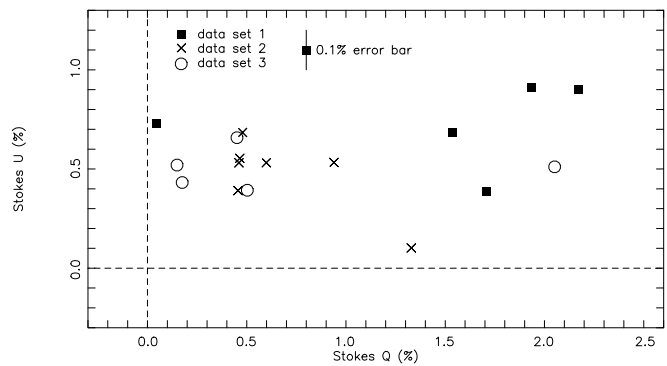}
\caption{Plot of the Stokes parameters observed for AA~Tau. The filled 
squares show data set 1, the crosses show data set 2, and the empty
circles show data set 3. The dashed-lines trace the lines where Q and
U are equal to zero, for reference. A representative 0.1\% error bar is
shown for comparison.}
\label{fig:QUplot}
\end{figure}

Although the data presented in \S~\ref{sec:obspol} were obtained in 3
different filters, they clearly show, globally but also within each
set, that the linear polarisation of AA~Tau is variable. This confirms
the variability analysis of M\'enard \& Bastien (1992).  The maximum
polarisation observed is P$_{\rm max}$ = 2.35\% in the blue and
P$_{\rm max}$ = 2.1\% in the red. Both values are similar, within
1-$\sigma$. The minimum value observed is P$_{\rm min}$ = 0.6\% in the
blue and P$_{\rm min} \sim$ 0.5\% in the red, also similar within
2-$\sigma$. The data is plotted in Fig.~\ref{fig:QUplot} where the
normalised Stokes parameters Q and U are plotted. They are related to
the usual polarisation, P, and position angle, $\theta$, by the
following relations:

\begin{equation}
{\rm Q} = {\rm P}\cos(2\theta), {\rm U} = {\rm P}\sin(2\theta).
\end{equation}

Clearly, the data are not distributed randomly in the QU-plane.  Each
measurement lies on an elongated strip. This strip
is bounded by $0.0\% \leq {\rm Q} \leq 2.2\%$ and $0.1\% \leq {\rm U}
\leq 0.9\%$.  It is elongated mostly in along the Q-axis.  It 
is important to note that the three data sets, taken in three
different filters at three different epochs, blend in smoothly within
the same strip. In \S~\ref{sec:discus} we will argue that this
behaviour of the polarisation in the QU-plane is intimately linked to
the geometry of the circumstellar environment of AA~Tau.

%----------------------------------------
\section{Discussion}
\label{sec:discus}

B99 presented an analysis of the variability of
AA~Tau based on extensive multi-site continuous spectro-photometric
monitoring of the star. From the light curve, they find a rotation
period of 8.2 days. The light curve is quite unusual
for a T Tauri star. It is roughly constant, in the bright state, with
quasi-periodic fadings of amplitude up to 1.4 magnitudes in
BVRI. Because of the lack of color variations during the fadings,
B99 interpreted the data in terms of occultations of
AA~Tau's photosphere by opaque circumstellar material. 

To produce the observed occultations, they proposed that the inner
edge of the accretion disk is bent, warped by a large scale dipole
tilted with respect to the stellar rotation axis and causing a
``piling-up'' of material preferentially on one side. This piling-up
results in non-axisymmetric vertical distribution of dust, the
occulting material. As the inner edge of the disk rotates, in
corotation with the star, it eclipses the photosphere partially every
8.2 days.

This magnetospheric accretion model has the ability to explain the
optical and near-infrared light curves and the veiling in a coherent
way. However, albeit interesting, the model raises a number of issues
that one must solve in order to push it further. This
section will address two of these issues. First, are the fadings seen
in the light curve really caused by extinction?  Second, is the
accretion disk tilted to a large inclination, next to edge-on, as
needed to produce eclipses?

Answering these questions is important to confirm the magnetospheric
accretion model for AA~Tau and understand the origin of the
peculiarities of its light curve and therefore study the applicability
of AA~Tau's model to other T Tauri stars.

\subsection{Evidence for circumstellar extinction}

AA~Tau's light variations have roughly neutral colors during the
fadings, sometime exhibiting a slight blueing at minimum light (B99).
This is very reminiscent of UXORs where deep minima are found, with
blueing for the deepest ones. The case for circumstellar extinction is
well documented for UXORs (Natta et al. 2000, and references
therein). Interestingly, the polarisation of UXORs increases when the
stars fade. This is because the relative fraction of scattered (hence
polarised) light increases as the photosphere becomes occulted by an
optically thick clump of circumstellar material.

\begin{figure}
\centering
\includegraphics[width=2.9cm, angle=-90.0]{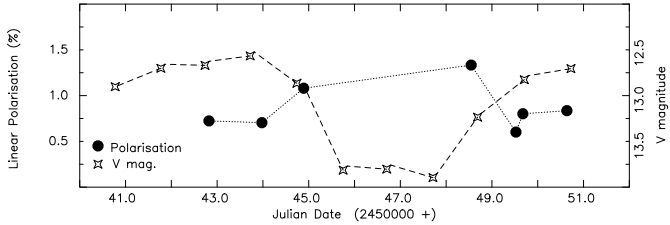}
\caption{Plot of the polarimetric and photometric variations of AA~Tau 
as a function of time during part of the 1995 campaign. Data are from
B99. {\bf Full circles \& dotted line:} V-band
linear polarisation of AA~Tau as a function of Julian date.{\bf Open
stars \& dashed line:} V-band photometry as a function of Julian
date.}
\label{fig:pvsv-95}
\end{figure}

B99 presented sparse linear polarimetry with their photometry. The
polarisation shows a hint of increasing when AA~Tau fades but the
measurements did not cover a full occultation, see
Fig.~\ref{fig:pvsv-95} where we reproduce the relevant
data. Therefore, no firm conclusions could be reached.

\begin{figure}
\centering
\includegraphics[width=4.1cm, angle=-90.0]{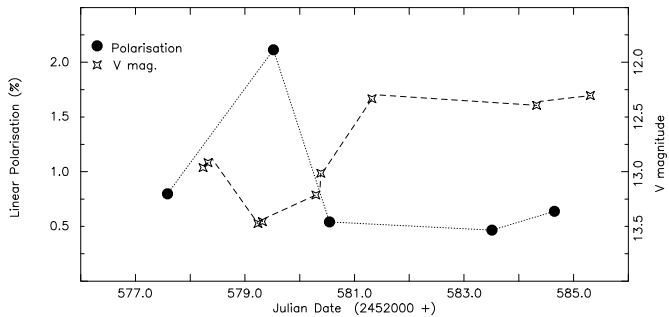}
\caption{Plot of the polarimetric and photometric variations of AA~Tau 
as a function of time. {\bf Full circles \& dotted line:} I-band
linear polarisation of AA~Tau as a function of Julian date.{\bf Open
stars \& dashed line:} V-band photometry as a function of Julian
date.}
\label{fig:pvsv-02}
\end{figure}

In Fig.~\ref{fig:pvsv-02} we plot the polarimetric data and the V-band
photometry of data set 3 (see Tab.~\ref{tab:poldata}) as a function of
time.  Clearly, close to JD=2452579.5 the polarisation rose from the
minimum level ($P_{\rm min}$) to the maximum value ($P_{\rm max}$),
and decreased back to the minimum level as the light curve went
through a minimum. The maximum polarisation is coincident
with the minimum brightness of AA~Tau. This is a strong indication
that extinction is responsible for the fading of AA~Tau and scattering
in the nearby circumstellar environment responsible for the
polarisation.
 
\subsection{Geometry of the occulting material}
\label{sec:inc}

In the model envisioned by B99, the maximum height of the warped disk
occurs at the same longitude as the photospheric hot spot, which is
also the longitude of the tilted magnetic pole. This is probably an
over simplification and further monitoring and numerical simulations
will help study this question in greater details.  For now, we will
consider that the disk is deformed and its increased vertical
extension is causing the observed occultations.  Irrespective of the
exact azimuthal location of this vertical extent though, simple
geometric arguments are sufficient to deduce the inclination to the
line-of-sight of the accretion disk from the temporal variations of
the polarisation curve, i.e, its shape in the QU-plane (see
Fig.~\ref{fig:QUplot}).

For a perfectly axisymmetric star-disk system, the ratio of direct
unpolarised light to scattered polarised light remains constant with
rotational phase. Photometry and polarimetry are constant in that
case. Including an asymmetry, e.g., a warped inner edge in the disk,
modifies the amount of scattered flux received by an observer in two
ways compared to the axisymmetric case. First, the total brightness of
the star-disk system will not remain the same. Second, it will also be
modulated as the system rotates.

For an observer viewing the system equator-on, the light distribution
will go through maximum amplitude variations as the warp in the disk
moves in and out of the line-of-sight, i.e., occulting the
photospheric disk in a variable way. The polarisation position angle
will remain constant because the disk is projected as a straight
line in the plane of the sky. In this case, only the total
brightness and polarisation level will change as the asymmetry moves
in and out of the line-of-sight.

On the contrary, an observer viewing the same non-axisymmetric system
pole-on would not detect a rotational modulation in the photometry
because the whole system is always viewed with the same aspect. Only a
rotation in the distribution of scattered light (i.e., in the
brightness map) and a rotation in the polarisation position angle
would occur, but without amplitude modulations. 

For AA~Tau, as shown in Fig.~\ref{fig:QUplot}, all the data points
fall within a narrow elongated strip, i.e., the polarisation level is
modulated but its position angle is constant. This argues
strongly in favor of a star-disk system seen at a large inclination,
almost edge-on.  The non-zero width of the strip where the data fall
in Fig.~\ref{fig:QUplot} can have several origins. On the one hand,
the system may not be perfectly edge-on and the polarisation curve not
a perfect straight line. On the other hand, the scattering geometry is
also likely to be more complex than assumed here.  Finally, long
term variability and non-zero error bars may contribute to increase
the observed width.

\subsection{Orientation of the accretion disk}
\label{sec:pa}

The arguments presented in \S~\ref{sec:inc} can also be used to
predict the position angle of the disk, the other angle defining its
complete orientation in 3D space. Indeed, we deduced that the
inclination is large. For simplicity, we will now assume that the
system is exactly edge-on and the polarisation curve in the QU-plane
(see Fig.~\ref{fig:QUplot}) is a straight line that is running
left-right. We will also assume that scattering on dust in responsible
for the polarisation. This is reasonable in comparison with other
accretion disks observed in T Tauri stars by scattered light in the
optical and near-infrared (e.g., M\'enard \& Bertout 1999 for a
review). In that case, the net polarisation is usually perpendicular
to the plane of the disk in the absence of a large envelope (e.g.,
Whitney \& Hartmann 1992). In Fig.~\ref{fig:QUplot}, the polarisation
curve runs parallel to the Q-axis. In the equatorial coordinate system
this is along the North-South direction, i.e., at a position angle
$\sim 0^o$. If the polarisation is indeed perpendicular the plane of
the disk, then we predict that AA~Tau's light scattering disk should
be oriented more or less along the East-West direction.

\subsubsection{On the variations of the observed position angle}

In general the Stokes parameters are the algebraic sum of two (or
more) contributions, intrinsic and interstellar:

\parbox{6cm}{\begin{eqnarray*}
{\rm Q_{observed} = Q_{intrinsic} + Q_{ISM}},\\
{\rm U_{observed} = U_{intrinsic} + U_{ISM}}.
\end{eqnarray*}}
\hfill
\parbox{1.0cm}{\begin{eqnarray}\end{eqnarray}}

The observed polarisation position angle is defined as $\theta
= \frac{1}{2} {\rm arctg}(\frac{\rm U_{obs}}{\rm Q_{obs}})$.
 
In the previous section we used the fact that the polarisation curve
is oriented along the Q-axis to predict that the disk will be oriented
East-West.  In reality the position angles we measure are variable and
in the range $2^o \leq \theta \leq 43^o$ instead, not quite at 0$^o$.
We show here that this behaviour is not intrinsic to the source.

In general, the interstellar polarisation is assumed to be
constant. From Eqs.~2, it is immediate to see that a constant ${\rm
Q_{ISM}}$ and ${\rm U_{ISM}}$ induce only a shift of the intrinsic
polarisation curve in the QU-plane. Any favored behaviour in the
polarisation curve will therefore trace the intrinsic properties of
the system. We need not care about the shift from zero of
AA~Tau's measured polarisation curve in the QU-plane to estimate the
orientation of its circumstellar environment. What matters is the
general orientation of this curve in Fig.~\ref{fig:QUplot}.

To go further, we can show that ${\rm U_{intrinsic}} \simeq 0$, hence
that $\theta_{intrinsic} \simeq 0$ for AA~Tau.  The observed
variations in position angles being due only to ${\rm U_{observed}}
\neq 0$ because ${\rm U_{ISM}} \neq 0$ and ${\rm Q_{obs}}$ being
variable.

To verify that the shift towards the +U axis of AA~Tau's polarisation
curve is caused only by interstellar polarisation, i.e., ${\rm
U_{observed} = U_{ISM}}$, we compare AA~Tau's polarisation with that
of other nearby T Tauri stars. Within 30 arcmin of AA~Tau, DN~Tau
(P=0.6\%, $\theta=34^o$), GI~Tau (P=0.8\%, $\theta=65^o$) , GK~Tau
(P=1.25\%, $\theta=42^o$), and V830~Tau (P=0.46\%, $\theta=53^o$) are
found for which we have polarisation measurements in the I-band, or
similar. Of these, V830~Tau is a weak-line T Tauri star. It is not
expected to have a significant amount of circumstellar dust, hence it
should not show a large intrinsic polarisation. We therefore attribute
an interstellar origin to its observed polarisation. Because it is
located near AA~Tau, it is reasonable to assume that AA~Tau also has
an interstellar polarisation of the same order, i.e., P$\sim$0.5\% at
$\sim50^o$. According to Eq.~(1), this polarisation has the following
Stokes parameters: ${\rm Q_{ISM}} = -0.09\%, {\rm U_{ISM}} =
+0.49.\%$.

The shift of AA~Tau's measurements along the Q-axis caused by
interstellar polarisation is minimal because ${\rm Q_{ISM}}$ is small,
as measured on V830~Tau. However, the observed shift along the U-axis
is of order 0.5\% and can be attributed entirely to the
interstellar component since ${\rm U_{ISM}} \sim +0.5\%$.  Removing
the interstellar contribution brings AA~Tau's curve down almost
exactly on the Q-axis.

As a consequence, we believe that the variations of the observed
linear polarisation position angle are not intrinsic to AA~Tau and do
not reflect a rapid change in the geometry of its circumstellar
environment. We predict that the light scattering disk is oriented in
the East-West direction.

\subsection{Comparison with other studies}

Dutrey et al. (1996) observed the dust thermal emission of AA Tau at
2.7mm with the IRAM interferometer. A flux of 13.2$\pm$1.8 mJy was
measured but the source remained unresolved in the $\sim$3$''$
beam. Recently, Kitamura et al. (2002) re-observed AA~Tau with the
Nobeyama millimeter array. At 2mm, they measured a flux of 22$\pm$3
and 13$\pm$2 mJy in compact (5'' beam) and extended (1'' beam)
configurations, respectively. AA~Tau is marginally resolved in their
images. They find a major axis of 1.34 $\pm$ 0.1 arcsec and a minor
axis of 0.61 $\pm$ 0.13 arcsec (with a synthesized beam size of 1.4
and 1.3 arcsec along the major and minor axis respectively).

The major axis of the outer disk is well resolved and Kitamura et
al. (2002) estimate a reliable position angle of $86^o\pm5^o$.  This
is in very good agreement with our suggestion that the inner disk runs
East-West (see \S~\ref{sec:pa}).

However, the minor axis of AA~Tau's disk is only marginally resolved
at radio wavelengths and the inclination, calculated as the ratio of
minor to major axis, is probably overestimated by their
measurements. They ran two different models to fit their data and
obtained inclination values in the range $i=25-28^o$, not compatible
with the model of B99 nor with the estimations presented in
\S~\ref{sec:inc}. Finer millimeter observations with more extended
arrays and at higher frequencies are needed to check these results and
solve this discrepancy.

%----------------------------------------
\section{Conclusions}

We presented optical linear polarisation data of the Classical T
Tauri star AA~Tau. The data were obtained at three different epochs
and show a remarkably stable behaviour: a low-level polarisation when
the star+disk system is bright and a significant increase in
polarisation when the system dims.

These results allow us to confirm that extinction by circumstellar
material is responsible for the quasi-periodic fading episodes
observed in the light curve. Furthermore, the distinctive shape of the
polarisation curve strongly supports the idea that the occulting
material orbiting AA~Tau is located in a disk viewed at a large
inclination, close to edge-on. Finally, the disk's major axis is
likely oriented in the East-West direction.

These results allow to confirm and refine the magnetospheric accretion
model put forward for AA~Tau. The orientation of AA~Tau's disk
provides the solution as to why its light curve is different from that
of most other classical T Tauri stars: the ``eclipses'' are a direct
consequence of its large tilt to the line-of-sight.

\begin{acknowledgements}

The authors gratefully acknowledge financial support from Universit\'e
Joseph Fourier, Grenoble (1994 BQR grant B 644 R1), the Laboratoire
d'Astrophysique de Grenoble, and the Observatoire Midi-Pyr\'en\'ees
for upgrade and maintenance of the {\sc sterenn} polarimeter.
Financial support for the observations at Pic-du-Midi was provided by
the Programme National de Physique Stellaire (PNPS) of CNRS/INSU,
France. This work was also supported by a NATO Science Program grant
(PST.CLG.976194).

\end{acknowledgements}

\end{document}